
\magnification=\magstep1
\hsize=13cm
\vsize=20cm
\overfullrule 0pt
\baselineskip=13pt plus1pt minus1pt
\lineskip=3.5pt plus1pt minus1pt
\lineskiplimit=3.5pt
\parskip=4pt plus1pt minus4pt

\def\negenspace{\kern-1.1em}



\newcount\secno
\secno=0
\newcount\susecno
\newcount\fmno\def\z{\global\advance\fmno by 1 \the\secno.
                       \the\susecno.\the\fmno}
\def\section#1{\global\advance\secno by 1
                \susecno=0 \fmno=0
                \centerline{\bf \the\secno. #1}\par}
\def\subsection#1{\medbreak\global\advance\susecno by 1
                  \fmno=0
       \noindent{\the\secno.\the\susecno. {\it #1}}\noindent}


\def\sqr#1#2{{\vcenter{\hrule height.#2pt\hbox{\vrule width.#2pt
height#1pt \kern#1pt \vrule width.#2pt}\hrule height.#2pt}}}


\newcount\refno
\refno=1
\def\y{\the\refno}
\def\myfoot#1{\footnote{$^{(\y)}$}{#1}
                 \advance\refno by 1}


\def\neq{\hbox{$\,$=\kern-6.5pt /$\,$}}





\newcount\secno
\secno=0
\newcount\fmno\def\z{\global\advance\fmno by 1 \the\secno.
                       \the\fmno}
\def\sectio#1{\medbreak\global\advance\secno by 1
                  \fmno=0
       \noindent{\the\secno. {\it #1}}\noindent}

\def\semidirect{\;{\rlap{$\supset$}\times}\;}




\input extdef
\magnification=\magstep1
\hsize 13cm
\vsize 20cm
\hfill{Preprint IMAFF 94/7}
\bigskip
\centerline{\bf{ORDINARY MATTER IN NONLINEAR AFFINE GAUGE
THEORIES}}
\centerline{\bf{OF GRAVITATION}}
\vskip 1.0cm
\centerline{by}
\vskip 1.0cm
\centerline{A. L¢pez--Pinto, A. Tiemblo and R. Tresguerres}
\vskip 1.0cm
\centerline{\it {IMAFF, Consejo Superior de Investigaciones Cient¡ficas,}}
\centerline{\it {Serrano 123, Madrid 28006, Spain}}
\vskip 1.5cm
\centerline{ABSTRACT}\bigskip
We present a general framework to include ordinary fermionic
matter in the metric--affine gauge theories of gravity. It is
based on a nonlinear gauge realization of the affine group,
with the Lorentz group as the classification subgroup of the
matter and gravitational fields.
\bigskip\bigskip

\sectio{\bf{Introduction }}\bigskip
Several generalizations of the gauge theory of gravitation
originally proposed by Utiyama, Sciama and Kibble$^{(1)}$
were developped in recent times$^{(2)}$. The metric--affine
gauge theories (MAGT's) propugnated by Hehl et al.$^{(3)}$ are
particularly interesting, since they provide a dynamical
foundation of the geometrical theory equiped with the most
general connection, namely with the general linear connection
$\Gamma _\alpha {}^\beta $. The affine group is the
transformation group of the geometry, thus being the main
candidate to describe things in the absence of matter.
According to Ne'eman and {\~{S}}ija{\~{c}}ki$^{(4)}$, the
adoption of the affine group as the dynamical group in a gauge
theory of gravitation could solve certain renormalizability and
unitarity problems in Quantum Gravity, with the help of the
additional degrees of freedom contained in the linear connection.

A main difficulty for the common acceptance of the MAGT's of
gravitation is the problem of including fermionic matter in this
general framework, since no finite linear half--integer
representations of the affine group exist. To avoid this
difficulty, infinite dimensional representations were suggested by
Ne'eman and {\~{S}}ija{\~{c}}ki$^{(4)}$. They demonstrated that
the groups SL(n,R) admite half--integer representations, called
{\it{manifields}}, suitable to describe fermionic matter, and
they proposed such a group as the classification group for
hadrons, which allows to study the corresponding
phaenomenological matter Lagrangians. But this does not solve
the problem of explaining the coupling of the affine
gravitational interactions to fundamental Dirac particles.

Nevertheless, the inclusion of ordinary matter in the affine
scheme becomes possible in the context of nonlinear realizations
of symmetry groups$^{(5)}$. The nonlinear coset realization,
initially proposed to treat internal symmetries, was soon
extended to spacetime symmetries. In fact, several attempts were
made to apply the nonlinear realizations to gravity$^{(6)}$.
In 1965, Ogievetskii and Polubarinov$^{(7)}$ found out a nonlinear
spinorial transformation law, which strongly suggested to explore
the possibility of a nonlinear realization of the affine group.
Borisov and Ogievetski$^{(8)}$ considered this possibility with
the Lorentz group as the classification subgroup. However, they
only took into account the global case, in conjunction with a
simultaneous nonlinear realization of the conformal group, since
they were interested in reproducing local diffeomorphisms from
the closure of the algebras of both groups. They did not
construct a gauge theory with degrees of freedom associated to
independent connections.\bigskip

\sectio{\bf{The affine and the Lorentz groups: geometry and
matter}} \bigskip
Einstein's aim was to construct a theory which guaranteed the
undistiguishability of the states of movement of arbitrary
observers. From the geometrical point of view, the most general
spacetime transformations are those of the affine group. The
affine tensor spaces$^{(9)}$ involve the vector basis
$e_\alpha :=\,e_\alpha {}^i\partial _i\,$, representing a
coordinate--independent local frame, and its dual 1--form basis
$\vartheta ^\beta\,$ called the coframe, defined by the
requirement that the interior product of both bases is
$e_\alpha\rfloor\vartheta ^\beta =\,\delta ^\beta _\alpha $. In
the gauge theories of the affine group, the coframe plays the
role of a nonlinear translational connection, invariant under
translations, see below. In addition, a linear connection
$\Gamma _\alpha {}^\beta\, $, necessary in view of the local covariance,
is in order. The group transformations act on the anholonomic indices.
In particular, the general linear group acts on the frame attached
to the observer, rotating, shearing and dilatating it.
Accordingly, the linear connections may be splitted up into an
antisymmetric, a symmetric traceless and a trace contribution.
In principle, these are, in the absence of matter, the dynamical
degrees of freedom involved.

However, the inclusion of matter drastically modifies the
situation. The role of the Lorentz group as the classification
group of the elementary particles oblies us to consider it as
the effective dynamical group. In so far, we described a pure
affine geometry without a metric. The metrization takes place
when we impose the condition that the speed of light remains a
constant in every local reference frame. This means that we have
to introduce an invariant Minkowski metric, so that the Lorentz
group becomes the only admissible transformation group to act on
the frames.

Matter fields must be placed on the linear representations of
the Lorentz group. Nevertheless, this does not mean that we have
to renounce to the original affine symmetry. One can realize the
affine group nonlinearly in such a way that the Lorentz group
plays the role of the classification subgroup, without imposing
constraints which reduce the number of degrees of freedom of the
original general linear connection. In fact, this is the most
natural way to match geometry and matter, preserving the simultaneous
validity of the spacetime groups characteristic for each one.

There is no an {\it{a priori}} reason to suppose that a given
physical symmetry must be realized linearly in nature. Indeed,
studying the general formulation of nonlinear realizations of
spacetime symmetry groups, the linear realizations manifest
themselves as a very special case. Let us mention some arguments
in favour of the nonlinear realization of the affine group. In
the first place, this group does not posses natural invariants,
as required to define the action of the theory. The conventional
approach to overcome this difficulty consists in distinguishing
two kinds of representations describing co- and contravariant
objects, transforming with $\Lambda $ and $\Lambda ^{-1}$
respectively, where $\Lambda $ is a general linear transformation.
This duplication of the representation space is related with the
inclusion of a metric tensor in the MAGT's, with ten additional,
dynamically (i.e. gauge--theoretically) not justified degrees of
freedom, playing a role similar as in the geometrical approach.
Invariants are then obtained saturating indices with the usual
rules. The dynamical role played by the metric tensor in the
MAGT's is obscure. An essential feature of the gauge theories
of internal groups is that the interactions are mediated
exclusively by connections, playing the role of gauge fields.
Nevertheless, as far as gravity is concerned, a metric tensor is
included by hand, involving dynamical degrees of freedom of
unknown origin.

We claim that a more natural solution to get invariants is to
project the action of the whole affine group into a subgroup
which admits such invariants. In particular, due also to the
dynamical reasons mentioned above, we choose the Lorentz group,
which has the Minkowski metric as a natural invariant. Thus, as
a consecuence of the nonlinear realization, it becomes possible
to construct invariants without increasing the number of degrees
of freedom of the theory, since the Minkowski metric is a
constant tensor. We will show below that the metric tensor of
the MAGT's can be obtained from our approach as a degenerate
case, by rearranging the hidden degrees of freedom contained in
the nonlinear realization, in such a way that the general linear
group becomes the classification subgroup.

The problem of the interpretion of the dynamical role played by
the metric tensor is closely related to that of the correct
treatment of the coframes. At this respect, a further argument
to consider nonlinear realizations of spacetime symmetry groups
is their success in explaining the link between the translational
connections and the coframes. This problem$^{(10)}$
is solved by means of a nonlinear realization of the
translations. The coframes turn out to be nonlinear
translational connections, with the right tensorial
transformation properties$^{(6)(11)}$.

In spite of the geometrical interpretation of the dynamical
objects, in particular of the translational connections, one has
to be careful in distinguishing the true dynamics from the misleading
manipulations allowed by the geometrical formalism, since
this may lead to the misunderstandings of the physical
meaning of both, the coframes and the metric tensor. Let us
consider an ilustrative example. As we have pointed out, in
a gauge theory the interactions are given by the connections.
In our nonlinear approach, the coframes depend on the
translational connection
${\buildrel (T)\over{\Gamma ^\alpha }}$ in the form
$\vartheta ^\alpha :=\,{\buildrel (T)\over{\Gamma ^\alpha}}
+D\xi ^\alpha \,$, see Ref.(11) and below. Thus, the coframe
cannot be brought into the value $\vartheta ^\alpha
=\,\delta ^\alpha _i d\,x^i$ in the whole space, unless in the
absence of interactions. Nevertheless, in practice one often
makes this choice in order to work with holonomic frames
(defined by the condition $d\,\vartheta ^\alpha =0\,$). This is
not rigourous dynamically, but geometrically it makes sense,
since it is equivalent to redefine the metric tensor. Indeed,
being the coframes 1--forms, we can write them down as
$\vartheta ^\alpha =\,e^\alpha {}_i d\,x^i\,$. Thus, one only
has to define the holonomic metric tensor
$g_{ij}:=\,o_{\alpha\beta }\,e^\alpha {}_i\,e^\beta {}_j\,$,
in order to realize the invariant line element alternatively
in the anholonomic and holonomic forms respectively as
$o_{\alpha\beta }\vartheta ^\alpha\otimes\vartheta ^\beta
=\,g_{ij}\,d\,x^i\otimes d\,x^j \,$, where $d\,x^i$ in the
r.h.s. plays the role of an holonomic coframe as discussed above.
This coframe does not posses the meaning of a gauge field any
more, since the dynamical information contained in the original
coframes with gauge theoretical meaning has been transfered to
the metric tensor.

A systematic approach to the nonlinear realization of groups is
given by the so called {\it{coset realizations}}$^{(5)}$, that
we are going to discuss in the following section.\bigskip

\sectio{\bf{Coset realizations of symmetry groups}}\bigskip
Let $G=\{g\}$ be a Lie group including a subgroup $H=\{h\}$
whose linear representations $\rho (h)$ are known, acting on
functions $\psi $ belonging to a representation space of $H$.
The elements of the quotient space $G/H$ are equivalence
classes of the form $gH=\{gh_1\,,gh_2\,...\,gh_n\}\,$,
and they constitute a complete partition of the group space.
We call the elements of the quotient space cosets to the left
(right) of $H$ with respect to $g$. Since we deal with Lie
groups, the elements of $G/H$ are labeled by continuous
parameters, say $\xi$. We represent the elements of $G/H$ by
means of the coset indicators $c(\xi )\,$, parametrized by
the coset parameters $\xi\,$, playing the role of
a kind of coordinates. The nonlinear coset realizations$^{(5)}$
are based on the action of the group on $G/H$, i.e., on a
partition of its own space. An arbitrary element $g\epsilon G$
acts on $G/H$ transforming a coset into another, that is
$$\eqalign{g:\, G/H&\rightarrow G/H\cr
c\,(\xi )&\rightarrow c\,(\xi ')\,,\cr }\eqno(\z)$$
according to the general law
$$g\,c\,(\xi\,)=\, c\,(\xi ')\, h\left(\xi\,, g\right)\,.\eqno(\z)$$
The elements $h\left(\xi\,,g\right)$ which appear in (3.2) belong
to the subgroup $H$, that we will call in the following the
{\it{classification subgroup}}, since the elements $g$ of the
whole group $G$ considered in (3.2) act nonlinearly on the
representation space of the classification subgroup $H$ according to
$$\psi '=\,\rho\left(h\left(\xi\,, g\right)\right)\psi\,,\eqno(\z)$$
where $\rho$, as mentioned above, is a linear representation
of $H$ in the space of the matter fields $\psi$. Therefore, the
action of the total group $G$ projects on the representations of
the subgroup $H$ through the dependence of $h\left(\xi\,,g\right)$
in (3.2) on the group element $g$, as given by eq.(3.3). The action of
the group is realized on the couples $\left(\xi\,,\psi\right)$.
It reduces to the usual linear action of $H$ when we take in
particular for $g$ in (3.2) an element of $H$.

In order to define a covariant differential transforming like
(3.3) under local transformations, we need a suitable nonlinear
connection. We define it as
$$\Gamma:=\, c^{-1}{\rm {\cal{D}}}c\,,\eqno(\z)$$
where the covariant differential on the coset space is defined as
$${\rm {\cal{D}}}c:=\,\left(d\,+\Omega\,\right)c\,,\eqno(\z)$$
with the ordinary linear connection $\Omega$ of the whole group G
transforming as
$$\Omega '=\, g\,\Omega \,g^{-1}+g\,d\,g^{-1}\,.\eqno(\z)$$
It is easy to prove that the nonlinear gauge field $\Gamma$
defined in (3.4) transforms as
$$\Gamma '=\,h \Gamma h^{-1} +h d\, h^{-1}\,,\eqno(\z)$$
thus allowing to define the nonlinear covariant differential
operator
$${\bf D}:=\,d\, +\Gamma\,.\eqno(\z)$$
One can read out from (3.7) that only the components of $\Gamma$
related to the generators of $H$ behave as true connections,
transforming inhomogeneously, whereas the components of $\Gamma$
over the generators associated with the cosets $c$ transform as
tensors with respect to the subgroup $H$ notwithstanding their
nature of connections.\bigskip

\sectio{\bf{Nonlinear gauge approach to the affine group}}\bigskip
As we have discussed above, ordinary matter fields such as
Dirac fields can only be included in an affine theory if the affine
group is realized nonlinearly, with the Lorentz group as the
classification subgroup. Thus, let us consider the affine group
$A(4\,,R)=\,GL(4\,,R)\semidirect R^{4}$ in $4$ dimensions,
defined as the semidirect product of the translations and the
general linear transformations, with the generators $\Lambda ^\alpha
{}_\beta $ of the linear transformations splitted up into the Lorentz
generators $L ^\alpha {}_\beta $, plus the remaining generators
$S ^\alpha {}_\beta $ of the symmetric linear transformations, as
$\Lambda ^\alpha {}_\beta =\,L ^\alpha {}_\beta
+ S ^\alpha {}_\beta $. In addition, we have the generators
$P_\alpha $ of the translations. The commutation relations read
$$\eqalign{\left[L_{\alpha\beta }\,,L_{\mu\nu }\right]=
-i\,&\left( o_{\alpha [\mu } L_{\nu ]\beta}
         - o_{\beta [\mu }L_{\nu ]\alpha }\right)\,,\cr
\left[L_{\alpha\beta }\,,S_{\mu\nu }\right]=
\,\,\,\,i\,&\left( o_{\alpha (\mu } S_{\nu )\beta}
         - o_{\beta (\mu }S_{\nu )\alpha }\right)\,,\cr
\left[S_{\alpha\beta }\,,S_{\mu\nu }\right]=
\,\,\,\,i\,&\left( o_{\alpha (\mu } L_{\nu )\beta}
         + o_{\beta (\mu }L_{\nu )\alpha }\right)\,,\cr
\left[L_{\alpha\beta }\,, P_\mu\right]\hskip0.10cm=
\,\,\,\,i\,&o_{\mu [\alpha }P_{\beta ]}\,,\cr
\left[S_{\alpha\beta }\,, P_\mu\right]\hskip0.10cm=
\,\,\,\,i\,&o_{\mu (\alpha }P_{\beta )}\,,\cr
\left[P_\alpha\,, P_\beta\right]\hskip0.30cm
=\,\,\,\,\,\,&\,0\,.\cr}\eqno(\z)$$
In order to realize the group action on the coset space
$A(4\,,R)/SO(1\,,3)$, we make use of the general formula (3.2)
which defines the nonlinear group action, choosing in particular
for the cosets the parametrization
$$c:=e^{-i\,\xi ^\alpha P_\alpha}e^{i\,h^{\mu\nu }S_{\mu\nu}}\,,\eqno(\z)$$
where $\xi ^\alpha$ and $h^{\mu\nu }$ are the coset parameters. The group
elements of the whole affine group $A(4\,,R)$ are parametrized as
$$g=\,  e^{i\,\epsilon ^{\alpha} P_\alpha}e^{i\,\alpha ^{\mu\nu }S_{\mu\nu}}
e^{i\,\beta ^{\mu\nu }L_{\mu\nu}}\,,\eqno(\z)$$
and those of the classification Lorentz subgroup are taken to be
$$h:=e^{i\,u ^{\mu\nu }L_{\mu\nu}}\,.\eqno(\z)$$
Other parametrizations are possible, leading to equivalent results.

In the Appendix {\bf A}, we show in some detail the calculations
which lead to the explicit form of the infinitesimal
transformations of the coset parameters. The variation of the
translational ones reads
$$\delta\xi ^{\alpha }=-\left(\alpha _\beta {}^\alpha
+\beta _\beta {}^\alpha \right)
\,\xi ^\beta -\epsilon ^\alpha\,,\eqno(\z)$$
thus assigning these parameters the role of coordinates,
transforming as usual under the affine group. On the other hand,
we obtain the variation of the matrix
$$r^{\alpha\beta }:=\, e^{h^{\alpha\beta }}\,,\eqno(\z)$$
constructed from the coset parameters $h^{\alpha\beta }$
associated to the symmetric affine transformations. In the
following, $r^{\alpha\beta }$ more than the coset parameters
itself, will play the fundamental role. Their variation reads
$$\delta r^{\alpha\beta }=\,\left(\alpha ^\alpha {}_\gamma
+\beta ^\alpha {}_\gamma\right)\, r^{\gamma\beta }+
u^\beta {}_\gamma\, r^{\gamma\alpha }\,.\eqno(\z)$$
Since $r^{\alpha\beta }$ is symmetric, the antisymmetric part
of (4.7) must vanish. From this symmetry condition, as shown in
Appendix {\bf B}, follows for the nonlinear Lorentz parameter
the expression
$$u^{\alpha\beta }=\,\beta ^{\alpha\beta }
-\alpha ^{\mu\nu }\tanh\left\{ {1\over 2}
\log\left[ r^\alpha {}_\mu\,\left( r^{-1}\right) ^\beta {}_\nu
\right]\right\}\,,\eqno(\z)$$
which shows how it consists of the linear Lorentz parameter
$\beta ^{\alpha\beta }$ plus a contribution depending on the
symmetric affine parameter $\alpha ^{\alpha\beta }$ {\it cum }
$r^{\alpha\beta }$. The action (3.3) of the full affine group
on arbitrary fields of a given representation space of the
Lorentz group reads infinitesimally
$$\delta\psi =\,i\, u^{\alpha\beta }\rho
\left(L_{\alpha\beta }\right)\psi\,,\eqno(\z)$$
being $u^{\alpha\beta }$ the nonlinear Lorentz parameter (4.8),
and $\rho\left(L_{\alpha\beta }\right)$ an arbitrary representation
of the Lorentz group. This equation is the key which allows to
include ordinary matter into an affine theory, since the action
of the symmetric linear transformations on the fields occurs
through the nonlinear Lorentz parameter, whereas the group
generators are the ordinary generators of the Lorentz group.
Thus, not only bosonic matter, but also Dirac fields, can be
considered as the sources of the gravitational potentials.

Now we will define the suitable connection for the nonlinear
gauge realization in two steps. We first introduce the
ordinary linear affine connection $\Omega $ in (3.5) as
$$\Omega :=-i\,{\buildrel (T)\over{\Gamma ^\alpha}} P_\alpha
         -i\,{\buildrel (GL)\over{\Gamma _\alpha {}^\beta }}
         \left( S^\alpha {}_\beta +L^\alpha {}_\beta \right)\,,\eqno(\z)$$
which includes the true translational potential
${\buildrel (T)\over{\Gamma ^\alpha}}\,$, and the $GL(4\,,R)$
connection ${\buildrel (GL)\over{\Gamma _\alpha {}^\beta }}$.
Their transformations (3.6) present the standard form
$$\delta {\buildrel (T)\over{\Gamma ^{\alpha}}} =\,
{\buildrel (GL)\over{D}}\epsilon ^\alpha
-\left(\alpha _\gamma {}^\alpha +\beta _\gamma {}^\alpha \right)
{\buildrel (T)\over{\Gamma ^{\gamma}}}\,,\eqno(\z)$$
and
$$\delta {\buildrel (GL)\over{\Gamma _\alpha {}^\beta }}=\,
{\buildrel (GL)\over{D}}\left(\alpha _\alpha {}^\beta
+\beta _\alpha {}^\beta \right)\,,\eqno(\z)$$
with ${\buildrel (GL)\over{D}}$ as the covariant differential
constructed with the $GL(4\,,R)$ connection. Eq.(4.12) shows that
both, the symmetric and the antisymmetric parts of the linear
connection ${\buildrel (GL)\over{\Gamma _\alpha {}^\beta }}$
transform as true connections, as expected. Making then use of
definition (3.4), we get the nonlinear connection
$$\Gamma :=\,c^{-1}\left(d\,+\Omega\,\right) c=-i\,\vartheta
^\alpha P_\alpha -i\,\Gamma _\alpha {}^\beta \left( S^\alpha {}_\beta
+L^\alpha {}_\beta\right)\,,\eqno(\z)$$
with the nonlinear translational connection $\vartheta ^\alpha $
and the nonlinear $GL(4\,,R)$ connection $\Gamma _\alpha
{}^\beta $ respectively defined as
$$\vartheta ^\alpha :=\,r_\beta {}^\alpha
\left( {\buildrel (T)\over{\Gamma ^\beta}}
+{\buildrel (GL)\over{D}}\xi ^\beta \right)\,,\eqno(\z)$$
and
$$\Gamma _\alpha {}^\beta :=\,\left( r^{-1}\right)_\alpha {}^\gamma
{\buildrel (GL)\over{\Gamma _\gamma {}^\lambda }}
\, r_\lambda {}^\beta -\left( r^{-1}\right)
_\alpha {}^\gamma \, d\,r_\gamma {}^\beta\,.\eqno(\z)$$
In the following, we will interpret (4.14) and (4.15) geometrically,
identifying $\vartheta ^\alpha $ with the coframe and
$\Gamma _\alpha {}^\beta $ with a {\it{geometrical}} connection
respectively. Eqs.(4.14,15) thus establish the correspondence between
the {\it{geometrical}} objects in the l.h.s. and the original
groupal objects in the r.h.s. According to (3.7), we find
that the coframe and the connection transform respectively as
$$\delta\vartheta ^{\alpha }
=-u_\beta {}^\alpha \vartheta ^\beta\,.\eqno(\z)$$
and
$$\delta \Gamma _\alpha {}^\beta =\,D\,u_\alpha {}^\beta \,,\eqno(\z)$$
with $D$ constructed with the nonlinear connection (4.15) itself.
Eqs.(4.16,17) show that the coframe transforms as a Lorentz vector,
with the nonlinear Lorentz parameter (4.8), and the connection
behaves as a Lorentz connection. In view of (4.13), this connection
is actually composed of two essentialy different parts, defined
on different elements of the Lie algebra of the affine
generators. Eq.(4.17) shows that only the antisymmetric part of
(4.15), defined on the Lorentz generators, behaves as a true
connection with respect to the Lorentz subgroup, whereas the
symmetric part is tensorial, transforming without any
inhomogeneous contribution, since $u^{\alpha\beta }$ is
antisymmetric. The geometrical meaning of the nonlinear
connection associated to the symmetric affine generators will
become clearer in terms of the Minkowski metric, as we are going
to show.

Since the nonlinear realization we are studying consists in a
projection of the action of the whole affine group on
its Lorentz subgroup, i.e. on the pseudoorthogonal group which,
by definition, leaves invariant the symmetric tensor
$$o_{\alpha\beta }:=\,diag(+\,-\,-\,-\,)\,,\eqno(\z)$$
this tensor appears automatically in the theory as a consequence
of the nonlinear treatment, being a natural invariant of
the Lorentz group. We interpret it geometrically in the common
way as the Minkowskian metric tensor. No degrees of freedom are
related to it. Its inverse reads
$$o^{\alpha\beta }:=\,diag(+\,-\,-\,-\,)\,,\eqno(\z)$$
and, as mentioned above, both are Lorentz invariant:
$$\delta o_{\alpha\beta }=\,0\quad\,,
\quad\delta o^{\alpha\beta }=\,0\,.\eqno(\z)$$
Defining the nonmetricity $Q_{\alpha\beta }$ as usual$^{(3)}$,
i.e. as minus the covariant differential of the metric tensor,
we find that it is proportional to the symmetric part of the
nonlinear $GL(4\,,R)$ connection, namely
$$Q_{\alpha\beta } :=-D o_{\alpha\beta }
=\,2\,\Gamma _{(\alpha\beta )}\,.\eqno(\z)$$
Thus, the geometrical meaning of the nonlinear connection
associated to the symmetric affine generators becomes apparent
as the nonmetricity, which behaves as a Lorentz tensor. In order
to show that, the transformation law (4.17) can be splitted up into
two parts, corresponding to the Lorentz connection
${\buildrel (Lor)\over{\Gamma _{\alpha\beta }}}
:=\,\Gamma _{[\alpha\beta ]}$, and to the nonmetricity
respectively. We get
$$\delta {\buildrel (Lor)\over{\Gamma _\alpha {}^\beta }}
=\,{\buildrel (Lor)\over{D}} u_\alpha {}^\beta \,,\eqno(\z)$$
and
$$\delta Q_{\alpha\beta }=\,2\,u\, _{(\alpha }{}^\gamma
Q_{\beta )\gamma }\,.\eqno(\z)$$
The tensorial character of the symmetric part of the connection
is a result of the nonlinear realization of the symmetric affine
transformations, but it is also evident from the definition (4.21)
of the nonmetricity, since $d o_{\alpha\beta }=0$.

Let us now find the natural field strengths of the theory from
the commutation of two covariant differentials operators (3.8).
This yields
$${\bf D}\wedge {\bf D} =-i\, T^\alpha P_\alpha
-i\,R _\alpha {}^\beta \left( S^\alpha {}_\beta +L^\alpha
{}_\beta \right)\,,\eqno(\z)$$
with the torsion $T^\alpha $ and the curvature $R _\alpha
{}^\beta $ respectively defined as
$$T^\alpha :=\,D\vartheta ^\alpha \,,\eqno(\z)$$
and
$$R _\alpha {}^\beta :=\, d\Gamma _\alpha {}^\beta
+\Gamma _\gamma {}^\beta\wedge\Gamma _\alpha {}^\gamma \,,\eqno(\z)$$
thus showing the groupal character of the torsion as the field
strength of the translations. The curvature has a symmetric and
an antisymmetric part, corresponding to the symmetric and the
antisymmetric (Lorentz) generators of the affine group
respectively. Both, the torsion and the curvature, transform as
Lorentz tensors, namely
$$\delta T^\alpha =-u_\beta {}^\alpha T^\beta \,,\eqno(\z)$$
and
$$\delta R _\alpha {}^\beta =\,u_\alpha {}^\gamma
R _\gamma {}^\beta -u_\gamma {}^\beta R _\alpha {}^\gamma
\,.\eqno(\z)$$

The geometrization of the gauge theory is completed by
introducing vector bases which represent the reference
frames. We interpreted the nonlinear translational connections
(4.14) as the coframes, which play the role of the 1--forms bases.
In terms of $\vartheta ^\alpha $, we define their dual vector
bases $e_\alpha$ by means of the interior product, as fulfiling
the general relation
$$e_\alpha\rfloor\vartheta ^\beta =\,\delta _\alpha ^\beta\,.\eqno(\z)$$
The vector bases thus transform as Lorentz vectors, namely
$$\delta e_\alpha =\,u_\alpha {}^\beta e_\beta\,.\eqno(\z)$$
Our nonlinear approach contains all the elements which appear in
the standard MAGT's, with the main difference of their
transformation properties. In particular, the metric tensor is
Lorentz invariant, i.e. it is fixed to be the Minkowskian
metric, which does not posses any degrees of freedom. This
seemengly makes a difference between the dynamical contents of
both theories, since the metric tensor in the MAGT's involves in
general ten degrees of freedom, but we will show below that
these can be factorized into the coframes and the connections.
\bigskip

\sectio{\bf Correspondence to the standard metric--affine case}
\bigskip
It is usually assumed in the context of
metric--affine gauge theories, that by means of a suitable
affine transformation one can allways choose the anholonomic
metric tensor in such a way that it becomes locally the
Minkowskian metric. This choice is reasonable since spacetime
is admitted to be locally Minkowskian, and it is also useful
for simplifying calculations when one looks for solutions of
the field equations of a given metric--affine Lagrangian. But
the deep meaning of the seemingly inocent choice
$g_{\alpha\beta }=o_{\alpha\beta}$ first becomes clear in
view of our results. When considered as globally valid, this
choice reveals itself as the key which allows
the transition from the standard metric--affine theories to the
nonlinear case with the Lorentz group as the classification
subgroup.

In principle, the nonlinear theory depends on the variables
$\xi ^\alpha $, $h^{\alpha\beta }$,
${\buildrel (T)\over{\Gamma ^\alpha}}$, and
${\buildrel (GL)\over{\Gamma _\alpha {}^\beta }}$, i.e., on
$4+10+16+64=\,94$ degrees of freedom. However, in the nonlinear
realization, these variables occur factorized in such a manner
that only the particular combinations $\vartheta ^\alpha $ and
$\Gamma _\alpha {}^\beta $, see (4.14,15), i.e. $16+64=\,80$
degrees of freedom, are relevant. Neither $\xi ^\alpha $ nor
$h^{\alpha\beta }$ appear explicitely as independent degrees of
freedom, since the symmetries to which they correspond are
realized nonlinearly.

We are interested in showing the correspondence between our
theory and the standard MAGT, in which only the translations are
realized nonlinearly, whereas the whole general linear group
acts linearly. In this case, only $\xi ^\alpha $ remains hidden,
whereas the ten degrees of freedom of $h^{\alpha\beta }$ must be
rearranged in a different way. In particular, they become
manifest through the metric tensor. Formally, the transition from the
nonlinear to the usual metric--affine objects resembles a finite
gauge transformation, with $r^{\alpha\beta }$ as the matrix of
the symmetric affine transformations. We represent the standard
metric--affine objects signed by a tilde. Then, the
corresponding coframes and connections are those of a nonlinear
realization of the affine group in which the role of the
classification group is played by the $GL(4\,,R)$
subgroup$^{(11)}$, namely
$$\tilde{\vartheta }^\alpha :=\,
{\buildrel (T)\over{\Gamma ^\alpha}}
+{\buildrel (GL)\over{D}}\xi ^\alpha
=\, \left( r^{-1}\right) _\beta {}^\alpha \vartheta ^\beta
\,,\eqno(\z)$$
and
$$\tilde{\Gamma }_\alpha {}^\beta :=
\,{\buildrel (GL)\over{\Gamma _\alpha {}^\beta }}
=\,r_\alpha {}^\gamma
\Gamma _\gamma {}^\lambda\left( r^{-1}\right)_\lambda {}^\beta
-r_\alpha {}^\gamma d\,\left( r^{-1}\right)_\gamma {}^\beta
\,.\eqno(\z)$$
In addition, the metric tensor, which is usually introduced by
hand in the affine scheme, and whose gauge--theoretical origin
is not manifest if one simply includes it in the nonlinear
realization with $GL(4\,,R)$ as the classification subgroup --in
which the coset parameters $h^{\alpha\beta }$ are absent--,
reveals itself as a particular factorization of the degrees of
freedom present in the nonlinear approach which takes the
Lorentz group as the classification subgroup. Actually, we have
$$\tilde{g}_{\alpha\beta }:=\,
r_\alpha {}^\mu r_\beta {}^\nu o_{\mu\nu }\quad\,,\quad
\tilde{g}^{\alpha\beta }:=\,
\left( r^{-1}\right) _\mu {}^\alpha
\left( r^{-1}\right) _\nu {}^\beta o^{\mu\nu }\,,\eqno(\z)$$
compare with Ref.(8). Thus, these objects are equivalent to
those of the framework of metric--affine theory. One only has to
fix the metric tensor to take globally the Minkowskian form, and
the degrees of freedom contained in it automatically reorganize
themselves in such a way that they reproduce our nonlinear
approach. As a consequence, observe that invariants like the
line element may be alternatively expressed in terms of the
nonlinear or metric--affine objects respectively, namely
$$ds^2=\,o_{\alpha\beta }\vartheta ^\alpha\otimes\vartheta ^\beta
=\,\tilde{g}_{\alpha\beta }\tilde{\vartheta }^\alpha
\otimes\tilde{\vartheta }^\beta \,,\eqno(\z)$$
where the transition from $o_{\alpha\beta }$ to
$\tilde{g}_{\alpha\beta }$ or {\it vice versa} takes place by
means of the suitable factorization of the coset parameters
associated to the symmetric affine transformations. The
gauge--theoretical origin of the metric tensor in the MAGT's is
thus explained. Moreover, the possibility of factorizing the
degrees of freedom of the metric into the coframes and the connections
explains why the field equations found by varying with respect
to the metric tensor in the framework of metric--affine theories
are not essential. Hehl et al.$^{(3)}$ showed that they are
redundant, when the field equations found from the coframes and
the linear connections are taken into account.

Making use of the transformation properties of
$r^{\alpha\beta }$, see (4.7), we find that the variations of
(5.1--3) are the expected ones, namely
$$\delta\tilde{\vartheta }^{\alpha }=
-\left(\alpha _\beta {}^\alpha +\beta _\beta {}^\alpha \right)
\tilde{\vartheta }^\beta \,,\eqno(\z)$$
$$\delta\tilde{\Gamma }_\alpha {}^\beta =\,\tilde{D}
\left(\alpha _\alpha {}^\beta +\beta _\alpha {}^\beta \right)\,,\eqno(\z)$$
and
$$\delta\tilde{g}_{\alpha\beta }
=\,2\left(\alpha _{(\alpha }{}^\gamma
+\beta _{(\alpha }{}^\gamma \right)\tilde{g}_{\beta )\gamma }
\quad\,,\quad\delta\tilde{g}^{\alpha\beta }
=-2\left(\alpha _\gamma {}^{(\alpha }
+\beta _\gamma {}^{(\alpha }\right) \tilde{g}^{\beta )\gamma }
\,.\eqno(\z)$$
The nonmetricity reads now
$$\tilde{Q}_{\alpha\beta } :=-{\buildrel (GL)\over{D}}
\tilde{g}_{\alpha\beta }=\,r_\alpha {}^\mu r_\beta {}^\nu
Q_{\mu\nu }\,,\eqno(\z)$$
and still transforms as a tensor, namely
$$\delta\tilde{Q}_{\alpha\beta }=2\left(\alpha _{(\alpha }{}^\gamma
+\beta _{(\alpha }{}^\gamma \right)\tilde{Q}_{\beta )\gamma }
\,,\eqno(\z)$$
but due to the fact that $d\,\tilde{g}_{\alpha\beta }\neq 0$,
the symmetric part of the connection is no more a tensor, but a
true connection. On the other hand, the metric--affine torsion and
curvature relate to the nonlinear ones through
$$\tilde{T}^\alpha :=\,{\buildrel (GL)\over{D}}
\tilde{\vartheta }^\alpha =\,
\left( r^{-1}\right) _\beta {}^\alpha T^\beta\,,\eqno(\z)$$
and
$$\tilde{R} _\alpha {}^\beta :=\,
d{\buildrel (GL)\over{\Gamma _\alpha {}^\beta }}
+{\buildrel (GL)\over{\Gamma _\gamma {}^\beta }}\wedge
{\buildrel (GL)\over{\Gamma _\alpha {}^\gamma }}
=\,r_\alpha {}^\mu \left( r^{-1}\right)_\nu {}^\beta R_\mu {}^\nu
\,,\eqno(\z)$$
and their variations read
$$\delta\tilde{T}^\alpha =
-\left(\alpha _\beta {}^\alpha +\beta _\beta {}^\alpha \right)
\tilde{T}^\beta\,,\eqno(\z)$$
and
$$\delta\tilde{R} _\alpha {}^\beta =
\left(\alpha _\alpha {}^\gamma +\beta _\alpha {}^\gamma \right)
\tilde{R}_\gamma {}^\beta
-\left(\alpha _\gamma {}^\beta +\beta _\gamma {}^\beta \right)
\tilde{R}_\alpha {}^\gamma\,.\eqno(\z)$$
The general law relating the variations of the fields $\psi $ of
our theory, see (4.9), and of those of the MAGT, can be written
in an abstract way as
$$\delta\left( e^{i\, h^{\mu\nu }S_{\mu\nu }}\psi\right) =\,
i\left(\alpha ^{\alpha\beta }S_{\alpha\beta }
+ \beta ^{\alpha\beta }L_{\alpha\beta }\right)
\left( e^{i\, h^{\mu\nu }S_{\mu\nu }}\psi\right)\,,\eqno(\z)$$
which does not present any problems as far as the regular
representations are concerned. How to interpret (5.14) in the case
of fields $\psi $ in a half--integer representation remains an
open question. The same comment holds for the relation between
the abstract differential operators, namely
$${\bf D}\psi =\,e^{-i\, h^{\mu\nu }S_{\mu\nu }}\tilde{{\bf D}}
\left( e^{i\, h^{\mu\nu }S_{\mu\nu }}\psi\right)\,.\eqno(\z)$$
In short, the standard MAGT's result from a particular
factorization of the ten degrees of freedom of the nonlinear
theory, which manifest themselves as the degrees of freedom of
the metric tensor.\bigskip\vfill\eject

\sectio{\bf{Conclusions}}\bigskip
The standard MAGT is a nonlinear realization of the affine
group in which only the translations are treated nonlinearly
--the nonlinear connections of the translations being identified
with the coframes--, whereas the general linear group plays the
role of the classification subgroup. In addition, a metric
tensor involving ten degrees of freedom is included. We have
proven that this theory, metric tensor included, can be obtained
from a more fundamental approach to gravity as a degenerate
case, where ordinary fermionic matter is absent.

In fact, the choice of the Lorentz group as the classification
subgroup, in a nonlinear realization, leads to a theory in which
ordinary Dirac fields can be included. The essential difference
between our approach and the standard metric--affine theories
consists in the transformation properties of the physical
fields, which behave as representations of the Lorentz group,
and in that the symmetric part of the connection is strictly
proportional to the nonmetricity, transforming as a tensor.
Nothing has to be changed in the standard metric--affine
formalism but the choice of the metric, and of course the
interpretation. The metric tensor is taken as the Minkowski
metric. Since it is invariant under Lorentz transformations,
it is globally fixed once and forever, so that space and time
are everywhere well defined as distinguished quantities. Being
a constant tensor, the metric has not a dynamical meaning any
more. The geometry is determined by the interactions, and in the
absence of gravity, the geometrical background simplifies to a
flat spacetime. But the effect of gravitation is, strictly
speaking, affine more than metric. The whole information about
the gravitational interactions is contained in the connections,
as in the gauge theories of internal groups.

Let us finally point out that, having the Lorentz group natural
invariants, this is also the case for the affine group when
realized nonlinearly with the Lorentz group as the
classification subgroup. In the context of MAGT's of
gravitation, McCrea$^{(12)}$ calculated the irreducible
decomposition of the curvature, the torsion and the nonmetricity
with respect to the Lorentz group. The utility of his work to
construct all possible invariants becomes evident in view of the
nonlinear realization presented here, since it enables us to
work with an affine geometrical formalism whose objects are
defined on representation spaces of the Lorentz group.
\bigskip\bigskip

\centerline{APPENDIX}\bigskip
\noindent {\bf A}.-- Let us show how the expressions (4.5) and (4.7)
were deduced (cfr. Ref.(8)). We apply the general formula (3.2)
defining the nonlinear action, namely
$$g\,c\,(\xi\,)=\, c\,(\xi ')\, h\left(\xi\,, g\right)\,,\eqno(A.1)$$
with the particular choices
$$g=\,  e^{i\,\epsilon ^{\alpha} P_\alpha}
e^{i\,\alpha ^{\mu\nu }S_{\mu\nu}}
e^{i\,\beta ^{\mu\nu }L_{\mu\nu}}\quad\,,\quad
c:=e^{-i\,\xi ^\alpha P_\alpha}e^{i\,h^{\mu\nu }S_{\mu\nu}}
\quad\,,\quad h:=e^{i\,u ^{\mu\nu }L_{\mu\nu}}\quad\,,\eqno(A.2)$$
as given in (4.2--4). The equation we have to solve then reads
$$e^{i\,\epsilon ^{\alpha} P_\alpha}e^{i\,\alpha ^{\mu\nu }S_{\mu\nu}}
e^{i\,\beta ^{\mu\nu }L_{\mu\nu}}e^{-i\,\xi ^\alpha P_\alpha}
e^{i\,h^{\mu\nu }S_{\mu\nu}} =\, e^{-i\,\xi ^{'\alpha} P_\alpha}
e^{i\,h^{'\mu\nu }S_{\mu\nu}}e^{i\,u ^{\mu\nu }L_{\mu\nu}}\,.\eqno(A.3)$$
We consider an affine transformation with infinitesimal group
parameters $\epsilon ^{\alpha }$, $\alpha ^{\mu\nu }$ and
$\beta ^{\mu\nu }$. Thus, the transformed coset parameters in
the r.h.s. of (A.3) reduce to $\xi ^{'\alpha} =\,\xi ^{\alpha}
+\delta\xi ^{\alpha}$ and $h^{'\mu\nu }=\,h^{\mu\nu }
+\delta h^{\mu\nu }$, and $u ^{\mu\nu }$ is also infinitesimal.
Consequently, eq.(A.3) reduces to
$$\eqalign{e^{i\,\xi ^\alpha P_\alpha }\bigl( 1+i\,\epsilon ^\alpha
P_\alpha +&i\,\alpha ^{\mu\nu }S_{\mu\nu }
+i\,\beta ^{\mu\nu }L_{\mu\nu }\bigr) e^{-i\,\xi ^\alpha P_\alpha }\cr
=&\,\left( 1-i\,\delta\xi ^\alpha P_\alpha \right)
e^{i\,( h^{\mu\nu }+\delta h^{\mu\nu }) S_{\mu\nu }}
\bigl( 1+i\,u ^{\mu\nu }L_{\mu\nu }\bigr)
e^{-i\,h^{\mu\nu }S_{\mu\nu }}\,.\cr }\eqno(A.4)$$
In order to simplify the calculations, we make use of the
generators of the whole linear transformations
$\Lambda ^\alpha {}_\beta :=\, L^\alpha {}_\beta
+S^\alpha {}_\beta\,$, in terms of which, the commutation
relations (4.1) reduce to
$$\eqalign{\left[\Lambda ^\alpha {}_\beta\,,\Lambda ^\mu {}_\nu\right]=&
\,i\,\left(\delta^\alpha _\nu \Lambda ^\mu {}_\beta
         -\delta^\mu _\beta \Lambda ^\alpha {}_\nu\right)\,,\cr
\left[\Lambda ^\alpha {}_\beta\,, P_\mu\right]\hskip0.15cm=&
\,i\,\delta^\alpha _\mu P_\beta\,,\cr
\left[P_\alpha\,, P_\beta\right]\hskip0.25cm
=&\,0\,.\cr}\eqno(A.5)$$
Making then use of the Campell--Hausdorff formula, we find the
algebraic relations
$$e^{i\xi ^\mu P_\mu }\omega _\alpha {}^\beta\Lambda ^\alpha {}_\beta
e^{-i\xi ^\mu P_\mu }=\,\omega _\alpha {}^\beta\Lambda ^\alpha {}_\beta
+\omega _\alpha {}^\beta\xi ^\alpha P_\beta \,,\eqno(A.6)$$
and
$$e^{i\zeta _\mu {}^\nu\Lambda ^\mu {}_\nu }
\kappa _\alpha {}^\beta\Lambda ^\alpha {}_\beta
e^{-i\zeta _\mu {}^\nu\Lambda ^\mu {}_\nu }
=\, e^{\zeta _\alpha {}^\mu }\kappa _\mu {}^\nu
e^{-\zeta _\nu {}^\beta }\Lambda ^\alpha {}_\beta
\,.\eqno(A.7)$$
Applying (A.6) to the l.h.s. of (A.4), it simplifies to
$$1+i\left[\epsilon ^{\alpha} +\left(\alpha _\beta{}^\alpha
+\beta _\beta{}^\alpha\right) \xi ^\beta\right] P_\alpha
+i\alpha ^{\mu\nu }S_{\mu\nu} +i\beta ^{\mu\nu }L_{\mu\nu}\,,
\eqno(A.8)$$
whereas in the r.h.s. we have to make use of the fact that
$$\eqalign{e^{i\,\left( h^{\mu\nu }+\delta h^{\mu\nu }\right)S_{\mu\nu}}
=&\, e^{i\,h^{\mu\nu }S_{\mu\nu}}\left( 1 +e^{-i\,h^{\mu\nu }S_{\mu\nu}}
\delta e^{i\,h^{\mu\nu }S_{\mu\nu}}\right)\cr
=&\, e^{i\,h^{\mu\nu }S_{\mu\nu}}\left[ 1
+i\, e^{-h^\alpha {}_\gamma }\delta e^{h^{\gamma\beta }}\left(
S_{\alpha\beta}+ L_{\alpha\beta}\right)\right]\,,\cr }\eqno(A.9)$$
in order to apply (A.7) particularized to the cases
$$e^{i\, h^{\mu\nu }S_{\mu\nu}}
\sigma ^{\alpha\beta}S_{\alpha\beta}
e^{-i\, h^{\mu\nu } S_{\mu\nu}}=\,
e^{h^\alpha {}_\mu }\sigma ^{\mu\nu }e^{-h_\nu {}^\beta }
\left( S_{\alpha\beta}+ L_{\alpha\beta}\right)\,,\eqno(A.10)$$
and
$$e^{i\, h^{\mu\nu }S_{\mu\nu}}
\tau ^{\alpha\beta}L_{\alpha\beta}
e^{-i\, h^{\mu\nu }S_{\mu\nu}}=\,
e^{h^\alpha {}_\mu }\tau ^{\mu\nu }e^{-h_\nu {}^\beta }
\left( S_{\alpha\beta}+ L_{\alpha\beta}\right)\,.\eqno(A.11)$$
Then, for $\xi ^{\alpha }$ and $r^{\alpha\beta }
:=\, e^{h^{\alpha\beta }}\,$, it follows
$$\delta\xi ^{\alpha }=-\left(\alpha _\beta {}^\alpha
+\beta _\beta {}^\alpha \right)
\,\xi ^\beta -\epsilon ^\alpha\,,\eqno(A.12)$$
and
$$\delta r^{\alpha\beta }=\,\left(\alpha ^\alpha {}_\gamma
+\beta ^\alpha {}_\gamma\right)\, r^{\gamma\beta }+
u^\beta {}_\gamma\, r^{\gamma\alpha }\,.\eqno(A.13)$$
Q.E.D.\bigskip

\noindent {\bf B}.--On the other hand, the nonlinear Lorentz
parameter (4.8) is calculated as follows. First we rewrite the
symmetry condition of (A.13), namely
$$u^{[\alpha }{}_\gamma\, r^{\beta ]\gamma }=\,
\left(\alpha ^{[\alpha }{}_\gamma +\beta ^{[\alpha }{}_\gamma
\right) r^{\beta ]\gamma }\,,\eqno(B.1)$$
in the form
$$\left( u^{\alpha\beta }-\beta ^{\alpha\beta }\right)
\left[ \delta ^\lambda _\alpha \delta ^\rho _\beta
+\left( r^{-1}\right) ^{[\lambda }{}_\alpha\, r^{\rho ]}{}_\beta
\right] =\,\alpha ^{\mu\nu }\left( r^{-1}\right) ^{[\lambda }{}_\mu\,
r^{\rho ]}{}_\nu\,,\eqno(B.2)$$
and we resolve it formally as
$$u^{\alpha\beta }=\,\beta ^{\alpha\beta }
+\alpha ^{\mu\nu }\left( r^{-1}\right) ^{[\lambda }{}_\mu\,
r^{\rho ]}{}_\nu\left[ \delta ^\lambda _\alpha \delta ^\rho _\beta
+\left( r^{-1}\right) ^{[\lambda }{}_\alpha\, r^{\rho ]}{}_\beta
\right] ^{-1}\,.\eqno(B.3)$$
Making then use of the fact that
$$\left( r^{-1}\right) ^{[\lambda }{}_\mu\, r^{\rho ]}{}_\nu =\,
{1\over 2}\left[ e^{\left( h^\rho {}_\nu -h^\lambda {}_\mu
\right)} - e^{-\left( h^\rho {}_\mu -h^\lambda {}_\nu
\right)}\right]\,,\eqno(B.4)$$
we recognize that
$$\eqalign{u^{\alpha\beta }=&\,\beta ^{\alpha\beta }
-\alpha ^{\mu\nu }\left\{ \sinh
\left( h^\lambda {}_\mu - h^\rho {}_\nu\right)
\left[ \delta ^\lambda _\alpha \delta ^\rho _\beta
+\cosh\left( h^\lambda {}_\alpha -h^\rho {}_\beta\right)
\right] ^{-1}\right\}\cr
=&\,\beta ^{\alpha\beta } -\alpha ^{\mu\nu }\tanh
\left( {{h^\alpha {}_\mu - h^\beta {}_\nu }\over 2}\right)\cr
=&\,\beta ^{\alpha\beta } -\alpha ^{\mu\nu }\tanh
\left\{{1\over 2}\log\left[ r^\alpha {}_\mu\,
\left( r^{-1}\right) ^\beta {}_\nu\right]\right\}\,.\cr }\eqno(B.5)$$
Q.E.D.\bigskip

\centerline {\bf Acknowledgement}\bigskip
\noindent We are very grateful to Dr. Jaime Julve for helpful
discussions.\bigskip\vfill\eject
\centerline{REFERENCES}\vskip1.0cm

\noindent\item{[1]} R. Utiyama,  {\it Phys. Rev.} {\bf 101} (1956) 1597

\item\qquad T. W. B. Kibble, {\it J. Math. Phys.} {\bf 2} (1961) 212

\item\qquad D. W. Sciama, {\it Rev. Mod. Phys.} {\bf 36} (1964) 463 and 1103

\noindent\item{[2]} A. Trautman, in {\it Differential Geometry},
Symposia Mathematica Vol. 12 (Academic Press, London, 1973), p. 139

\item\qquad A. G. Agnese and P. Calvini, {\it Phys. Rev.} {\bf
D 12} (1975) 3800 and 3804

\item\qquad F. W. Hehl, P. von der Heyde, G. D. Kerlick and J. M. Nester,
{\it Rev. Mod. Phys.} {\bf 48} (1976) 393

\item\qquad P. von der Heyde, {\it Phys. Lett.} {\bf 58
A} (1976) 141

\item\qquad E.A. Ivanov and J. Niederle, {\it Phys. Rev.} {\bf D25}
(1982) 976 and 988

\item\qquad D. Ivanenko and G.A. Sardanashvily, {\it
Phys. Rep.} {\bf 94} (1983) 1

\item\qquad E. A. Lord, {\it J. Math. Phys.} {\bf 27} (1986) 2415 and 3051

\noindent\item{[3]} F. W. Hehl, J. D. McCrea,  E. W. Mielke, and Y. Ne'eman
{\it Found. Phys.} {\bf 19} (1989) 1075

\item\qquad R. D. Hecht and F. W. Hehl, {\it {Proc. 9th Italian
Conf. G.R. and Grav. Phys.}}, Capri (Napoli). R. Cianci
et al.(eds.) (World Scientific, Singapore, 1991) p. 246

\item\qquad F.W. Hehl, J.D. McCrea, E.W. Mielke, and Y. Ne'eman, {\it
Physics Reports}, to be published.

\noindent\item{[4]} Y. Ne'eman and Dj. {\~{S}}ija{\~{c}}ki,
{\it Phys. Lett.} {\bf  B200} (1988) 489

Y. Ne'eman and Dj. {\~{S}}ija{\~{c}}ki, {\it Phys. Rev.} {\bf  D 37}
(1988) 3267

\noindent\item{[5]} S. Coleman, J. Wess and B. Zumino, {\it Phys. Rev.}
{\bf 117} (1969) 2239

\item\qquad C.G. Callan, S. Coleman, J. Wess and B. Zumino, {\it Phys. Rev.}
{\bf 117} (1969) 2247

\item\qquad E. W. Mielke, {\it Fortschr. Phys.} {\bf 25} (1977)
401, and references therein

\item\qquad S. Coleman, {\it Aspects of Symmetry}. Cambridge
University Press, Cambridge (1985)

\noindent\item{[6]} A. Salam and J. Strathdee, {\it Phys. Rev.}
{\bf 184} (1969) 1750 and 1760

\item\qquad C.J. Isham, A. Salam and J. Strathdee, {\it Ann. of Phys.}
{\bf 62} (1971) 98

\item\qquad L.N. Chang and F. Mansouri, {\it Phys.
Lett.} {\bf 78 B} (1979) 274, and {\it Phys. Rev.} {\bf D 17} (1978) 3168

\item\qquad K.S. Stelle and P.C. West, {\it Phys. Rev.}
{\bf D 21} (1980) 1466

\item\qquad E.A. Lord, {\it{Gen. Rel. Grav.}} {\bf 19} (1987) 983, and
{\it J. Math. Phys.} {\bf 29} (1988) 258

\noindent\item{[7]} A.B. Borisov and I.V. Polubarinov, {\it Zh.
ksp. Theor. Fiz.} {\bf 48} (1965) 1625, and V. Ogievetsky and
I. Polubarinov, {\it Ann. Phys.} (NY) {\bf 35} (1965) 167

\noindent\item{[8]} A.B. Borisov and V.I. Ogievetskii, {\it Theor.
Mat. Fiz.} {\bf 21} (1974) 329

\noindent\item{[9]} E. Cartan, {\it Sur les vari't's … connexion
affine et la th'orie de la relativit' g'n'ralis'e}, Ouvres
completes, Editions du C.N.R.S. (1984), Partie III. 1, pgs. 659
and 921

\noindent\item{[10]} K. Hayashi and T. Nakano, {\it Prog. Theor. Phys}
{\bf 38} (1967) 491

\item\qquad K. Hayashi and T. Shirafuji, {\it Prog. Theor. Phys}
{\bf 64} (1980) 866 and {\bf 80} (1988) 711

\item\qquad J. Hennig and J. Nitsch, {\it Gen. Rel.
Grav.} {\bf 13} (1981) 947

\item\qquad H.R. Pagels, {\it{Phys. Rev.}} {\bf D 29} (1984) 1690

\item\qquad T. Kawai, {\it{Gen. Rel. Grav.}} {\bf 18} (1986) 995

\item\qquad G. Grignani and G. Nardelli, {\it Phys. Rev.}
{\bf D 45} (1992) 2719

\item\qquad E. W. Mielke, J.D. McCrea, Y. Ne'eman and F.W. Hehl
{\it Phys. Rev.} {\bf D 48} (1993) 673, and references therein

\noindent\item{[11]} J. Julve, A. L¢pez--Pinto, A. Tiemblo and
R. Tresguerres, {\it{ Nonlinear gauge realizations of spacetime
symmetries including translations}}, (1994), to be published

\noindent\item{[12]} J. D. McCrea, {\it Class. Quantum Grav.}
{\bf 9} (1992) 553
\vfill\eject

\bye